\documentclass[
reprint,   
 amsmath,amssymb,
 aps,
]{revtex4-2}

\usepackage{graphicx}
\usepackage{dcolumn}
\usepackage{bm}
\usepackage{siunitx}
\usepackage{inputenc}
\usepackage{miller}
\DeclareSIUnit\nanosecond{ns} 
\DeclareSIUnit\megahertz{MHz}
\DeclareSIUnit\watt{W} 
\DeclareSIUnit\milliwatt{mW} 
\DeclareSIUnit\nanometer{nm} 
\DeclareSIUnit\picoseconds{ps} 
\DeclareSIUnit\gigahertz{GHz}
\DeclareSIUnit\ppm{ppm}
\DeclareSIUnit\ppb{ppb}

\begin{document}

\title{Wavelength Dependence of the Electrical and Optical Readout \\ of NV Centers in Diamond}

\author{Lina M. Todenhagen$^{1,2}$}
\email{lina.todenhagen@wsi.tum.de}
\author{Martin S. Brandt$^{1,2,3}$}
\affiliation{$^1$Walter Schottky Institut, Technische Universität München, Am Coulombwall 4, 85748 Garching, Germany}
\affiliation{$^2$Physik-Department, School of Natural Sciences, Technische Universität München, James-Franck-Straße 1, 85748 Garching, Germany}
\affiliation{$^3$Munich Center for Quantum Science and Technology (MCQST),\\ Schellingstraße 4, 80799 München, Germany}

\date{\today}

\begin{abstract}
We study the contrast for electrical and optical readout of NV centers in diamond in dependence of the optical excitation wavelength using different excitation schemes. While the optically detected magnetic resonance (ODMR) works efficiently between 480 and $\SI{580}{\nanometer}$, electrically detected magnetic resonance (EDMR) shows a strong dependence on the excitation dynamics. The highest, electrically detected contrast of $\SI{-23}{\percent}$ is achieved by resonantly exciting the zero-phonon line of the neutral charge state of NV at $\SI{575}{\nanometer}$. EDMR is also enhanced at $\SI{521}{\nanometer}$, possibly due to a further excited state of NV$^-$.
\end{abstract}

\maketitle
The nitrogen-vacancy (NV) center in diamond is one of the quantum systems frequently used in practical applications, especially for quantum sensing \cite{PaperWL-QuantumSesingMaterial, PaperWL-Friedemann}, communication \cite{PaperWL_NVEntanglement}, and computing \cite{PaperWL_QuantumComputingMeijer}. However, the widely applied optically detected magnetic resonance (ODMR) requires a complex setup and is often limited by the inefficient outcoupling of photons \cite{PaperWL-RefIndexB, PaperWL-RefractiveIndex1}. Alternatively, one can directly observe NV centers via electrically or photoelectrically detected magnetic resonance (EDMR or PDMR, respectively), as demonstrated for the first time by Bourgeois et al.~in 2015 \cite{PaperWL-B-Nature2015}. Meanwhile, also full spin readout \cite{PaperWL-Hrubesch}, single spin sensitivity \cite{PaperWL-Siyushev-SingleNV} and the detection of nuclear spins \cite{PaperWl-Morishita-nuclear} have been demonstrated using EDMR techniques. The implementation in all-diamond p-i-n devices \cite{PaperWL-PINDioden} underlines the commercial use-case of the electrical readout. 
Here, we identify the optimal excitation wavelength $\lambda_{\text{ex}}$ to maximize the contrast in EDMR and ODMR under different optical excitation conditions, realized by a pulsed supercontinuum source and a tunable continuous wave (cw) laser. While both readout methods yield broad contrast maxima between 480 and $\SI{580}{\nanometer}$, we additionally observe two resonant enhancements in the EDMR around 521 and $\SI{575}{\nanometer}$, reflected by corresponding minima in the ODMR. We attribute the first to an enhanced ionization of NV$^-$ via a resonant state in the conduction band \cite{PaperWL-Beha, PaperWL-2LevelExcitation-S} and the second is the zero-phonon line (ZPL) of NV$^0$ that allows for efficient recharging to NV$^-$ \cite{PaperWL-AugerProzess-ZPLExcitation}. Upon resonant excitation at $\SI{575}{\nanometer}$, we find the highest EDMR contrast of $\SI{-23}{\percent}$ for the IIa diamond material investigated here. Our results demonstrate the importance of systematic wavelength- and excitation dynamic-dependent experiments to disentangle the complex ionization and recharging cycle of NV centers in EDMR. Also, identifying the optimum excitation wavelength guides the further design of integrated sensors based on electrical readout. \par
\begin{figure*}
\includegraphics[width=\textwidth]{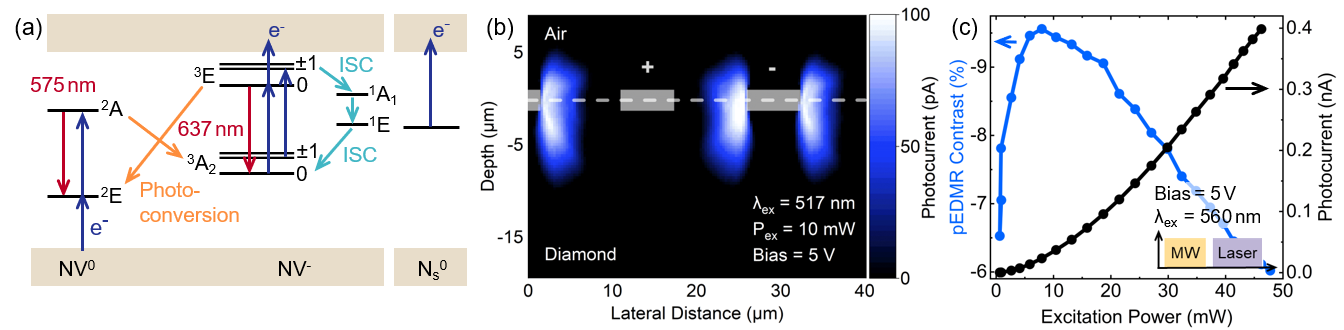}
\caption{\label{Pic1} (a) Excitation (dark blue arrows) of photoelectrically active defects in diamond. For $\lambda_{\text{ex}} \geq \SI{480}{\nanometer}$, ionization of NV$^-$ and recharging from NV$^0$ (orange arrows) require two photons each. Non-radiative intersystem crossing to the $^1$A$_1$ singlet (light blue arrows) is more likely to occur from the $m_s=\pm1$ levels of $^3$E, such that PL (red arrows) and PC are higher for $m_s=0$. N$_{\text{s}}^0$ can be ionized by a single photon and contributes to the background current. (b) Spatial photocurrent map recorded by scanning the sample with the excitation laser. Photocurrent originates from the vicinity of the negatively biased electrode (gray). (c) pEDMR contrast (blue) is measured for each wavelength as a function of $P_{\text{ex}}$ and only the highest values (here at $\SI{8}{\milliwatt}$ for $\lambda_{\text{ex}} = \SI{560}{\nanometer}$) contribute to Fig.~\ref{Pic2} and \ref{Pic3}. The same applies to pODMR. The absolute photocurrent under pulsed excitation (black) shows a superlinear power dependence, as expected for a two-step ionization of the NV center.}
\end{figure*}
The standard ODMR protocol is based on the spin-dependent luminescence of NV$^-$ and starts from the triplet ground state $^3$A$_2$, using $\lambda_{\text{ex}} \leq \SI{637}{\nanometer}$ for spin-conserving excitation to the $^3$E triplet (Fig.~\ref{Pic1}a). The return to $^3$A$_2$ can then occur either directly by emitting a photon or non-radiatively via intersystem crossing (ISC) to the $^1$A$_1$ singlet. As the ISC is more likely for the $m_s = \pm 1$ sublevels, the photoluminescence (PL) and the lifetime of the excited triplet become spin-dependent. Consequently, also the photoionization probability of $^3$E is spin-dependent, leading to a higher photocurrent (PC) from the $m_s = 0$ sublevel \cite{PaperWL-BReview}. Therefore, the EDMR of NV centers originates from a two-step process with $^3$E as intermediate level, which is reflected in a quadratic dependence of PC on the optical excitation power (Fig.~\ref{Pic1}c) \cite{LowTempNVPhotoIonization-Siyushev, PaperWL-B-Nature2015}. Direct ionization from the $^3$A$_2$ ground state, on the other hand, is a single-photon process with an experimentally reported threshold wavelength of $\SI{455}{\nanometer}$ \cite{PaperWL-DualBeamB} or $\SI{475}{\nanometer}$ \cite{PaperWL-Aslam} and contributes to the spin-independent background current, together with the ionization of nitrogen donors N$_{\text{s}}^0$ (also a single-photon process, threshold wavelength $\sim \SI{563}{\nanometer}$ \cite{PaperWL-RosaNesladek_PhotocurrentSpectroscopyLowE}). In the purely photoinduced NV charge state conversion cycle considered here, electrons needed for the recharging are excited directly from the valence band \cite{LowTempNVPhotoIonization-Siyushev, PaperWL-Razinkovas2}, which requires two photons of $\SI{445}{\nanometer} \leq \lambda_{\text{ex}} \leq \SI{575}{\nanometer}$ \cite{PaperWL-Aslam}. As the whole conversion cycle depends on the ionization probability of $^3$E, also the hole current generated in the recharging becomes spin-dependent.\par
All measurements were carried out at room temperature on type IIa diamond grown by chemical vapor deposition (CVD, SC Plate $\hkl<1 0 0>$, element six), which contains about $\SI{170}{\ppb}$ of N$_{\text{s}}^0$, $\SI{1}{\ppb}$ of NV and small amounts of H3 and NVH. For PC measurements ($\SI{5}{V}$ bias voltage), we deposit interdigitated Ti/Pt/Au electrodes of $\SI{10}{\micro\meter}$ distance on the diamond surface. A perpendicular bonding wire serves as microwave antenna to induce transitions between the $m_s = 0, \pm 1$ levels of $^3$A$_2$ at $\SI{2.87}{\gigahertz}$.
PC measurements are performed analogously to \cite{PaperWL-Hrubesch}, while we simultaneously detect the PL with a single-photon avalanche photodiode, protected by a $\SI{645}{\nanometer}$ longpass filter that also removes about $\SI{65}{\percent}$ of the PL from NV$^0$. For illumination, we use two laser sources with different spectral and temporal output characteristics. The first one is a pulsed supercontinuum source (SuperK FIU-15, NKT) with a repetition rate of $\SI{78}{\megahertz}$ and $\SI{50}{\picoseconds}$ pulse length, combined with an LLTF Contrast filter (excitation width $\Delta \lambda_{\text{ex}} \leq \SI{3.0}{\nanometer}$). For this excitation scheme, we use the term quasi-continuous wave (qcw) to highlight the short-pulsed nature of the optical illumination. The second laser source is a tunable continuous-wave laser (C-Wave VIS, HÜBNER Photonics) with $\Delta \lambda_{\text{ex}} \approx \SI{10}{\megahertz}$ in combination with an acousto-optic modulator to implement pulsed measurements. A 100x objective focuses the laser light on a spot that contains about 150 NV centers. We record spatial PC maps as shown in Fig.~\ref{Pic1}b by scanning the spot over the diamond, where PC is generated only in close proximity to the negatively-biased electrode as expected in a metal-semiconductor-metal photoconductor \cite{SZE-InBook}. For each excitation wavelength, an XYZ-piezo scanner automatically repositions the diamond to the PC maximum. 
The contrast was measured without applying a magnetic field and is calculated from $C = (I_{\text{res}} - I_{\text{off}})/I_{\text{off}}$, with $I_{\text{res}}$ and $I_{\text{off}}$ being the signal intensity upon resonant and off-resonant microwave irradiation, respectively.\par 
\begin{figure}
\includegraphics[width = \linewidth]{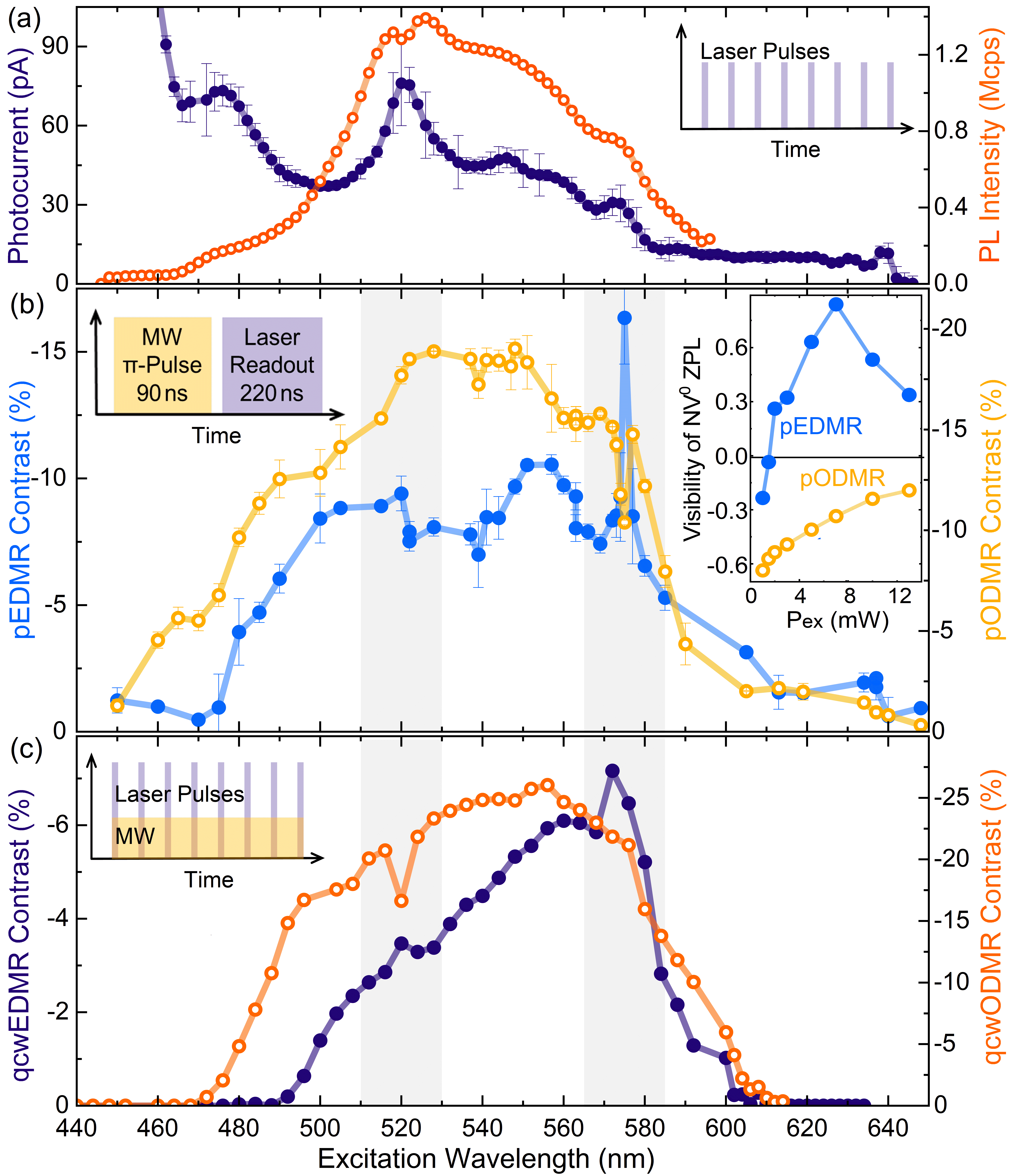}
\caption{\label{Pic2} (a) Photoluminescence and photocurrent excitation spectra under quasi-continuous wave excitation. PL and PC are both excited efficiently between $\SI{500}{\nanometer}$ and $\SI{575}{\nanometer}$. The background PC shows an additional increase towards shorter wavelengths. Dominant features in the PCE spectrum are the ZPL of NV$^0$ ($\SI{575}{\nanometer}$) and the peak at $\SI{521}{\nanometer}$, attributed to an electronic level in the conduction band. (b) Pulsed EDMR and ODMR for optimized $P_{\text{ex}}$ (cf. Fig.~\ref{Pic1}c) both yield high contrast between 480 and $\SI{580}{\nanometer}$ with a pronounced pEDMR maximum of $C\approx \SI{-17}{\percent}$ at $\lambda_{\text{ex}}=\SI{575}{\nanometer}$, accompanied by a minimum of $\SI{-10}{\percent}$ in the pODMR contrast. The strong $P_{\text{ex}}$ dependence of this feature is shown in terms of the ZPL visibility in the inset. (c) qcwEDMR and qcwODMR contrast measured with the supercontinuum source. Due to an increased contribution from single-photon ionization for shorter excitation wavelengths, the qcwEDMR contrast appears to be red-shifted. Besides the ZPL of NV$^0$, we find a second peak (dip) in EDMR (ODMR) around $\SI{521}{\nanometer}$. In all panels, insets depict the respective excitation schemes.}
\end{figure}
To understand the individual excitation processes leading to the formation of EDMR and ODMR contrast, we first measure photoluminescence (PLE) and photocurrent excitation spectra (PCE). For this, we use the supercontinuum source at constant $P_{\text{ex}} =\SI{1}{\milli\watt}$ (Fig.~\ref{Pic2}a), but note that the high peak power during the pulses is outside the range of linear response, where most driven transitions already saturate. For pulsed measurements (pEDMR and pODMR) with the tunable cw laser, we apply a pulse sequence consisting of a microwave $\pi$-pulse to manipulate the spin state, followed by a $\SI{220}{\nano\second}$ laser pulse for readout and re-initialization (inset in Fig.~\ref{Pic2}b). To obtain the maximum pODMR and pEDMR contrast, the $P_{\text{ex}}$ used was optimized for each wavelength and readout method separately as illustrated in Fig.~\ref{Pic1}c. A direct comparison of pODMR and pEDMR in Fig.~\ref{Pic2}b to the excitation spectra reveals, that the contrast is ultimately limited by the same processes as PLE. We discuss these findings by distinguishing three spectral regions in which we can drive different electronic transitions of NV$^-$, NV$^0$ and N$_{\text{s}}^0$.\par
In the first region below $\SI{480}{\nanometer}$, optical excitation directly ionizes NV$^-$ from the $^3$A$_2$ ground state, which effectively pumps the neutral charge state \cite{PaperWL_Waldherr} and results in the high PC and low PL intensities in Fig.~\ref{Pic2}a. The onset of PL generation occurs in two steps at $\lambda_{\text{ex}} = \SI{445}{\nanometer}$ and $\SI{470}{\nanometer}$, which we associate with the transition from a single to a two-photon process for recharging and ionizing NV$^-$, respectively \cite{PaperWL-Beha, PaperWL-Aslam}. However, direct ionization of NV$^-$ prevents the effective population of the $^3$E level where the actually spin-dependent ISC to the $^1\text{A}$ singlet level takes place. Therefore, PC and PL both depend only weakly on the NV$^-$ spin state and we observe low EDMR and ODMR contrast. EDMR is further suppressed, as direct ionization additionally increases the spin-independent background current from NV and also the ionization cross section of N$_{\text{s}}^0$ increases towards shorter $\lambda_{\text{ex}}$ \cite{PaperWL-Razinkovas2,PaperWL-RosaNesladek_PhotocurrentSpectroscopyLowE}.\par
Upon optical excitation in the second region between 480 and $\SI{575}{\nanometer}$, the ionization of NV$^-$ must now occur via the excited triplet state $^3$E, and the charge state interconversion between NV$^-$ and NV$^0$ is a two-step process in both directions. This not only leads to high PLE rates in Fig.~\ref{Pic2}a, but also causes a significant increase in pEDMR and pODMR contrast. Both contrasts remain high over the entire regime of two-photon ionization, ranging from $-12$ to $\SI{-18}{\percent}$ in the optical readout and from $-8$ to $\SI{-12}{\percent}$ in the electrical readout. When exciting at exactly $\SI{575}{\nanometer}$, corresponding to the ZPL of NV$^0$, we find the highest electrical contrast of almost $\SI{-17}{\percent}$ for this sample. Simultaneously, pODMR exhibits an equally sharp dip.\par
In the third region beyond $\SI{575}{\nanometer}$, we find a drastic decrease in both EDMR and ODMR contrast, as well as in PLE and PCE efficiency. This results from breaking the NV charge state conversion cycle, as the transition between NV$^0$ ground and excited state can no longer be driven. Instead, we effectively pump the NV center into its now dark, neutral charge state.\par
\begin{figure}
\includegraphics[width = \linewidth]{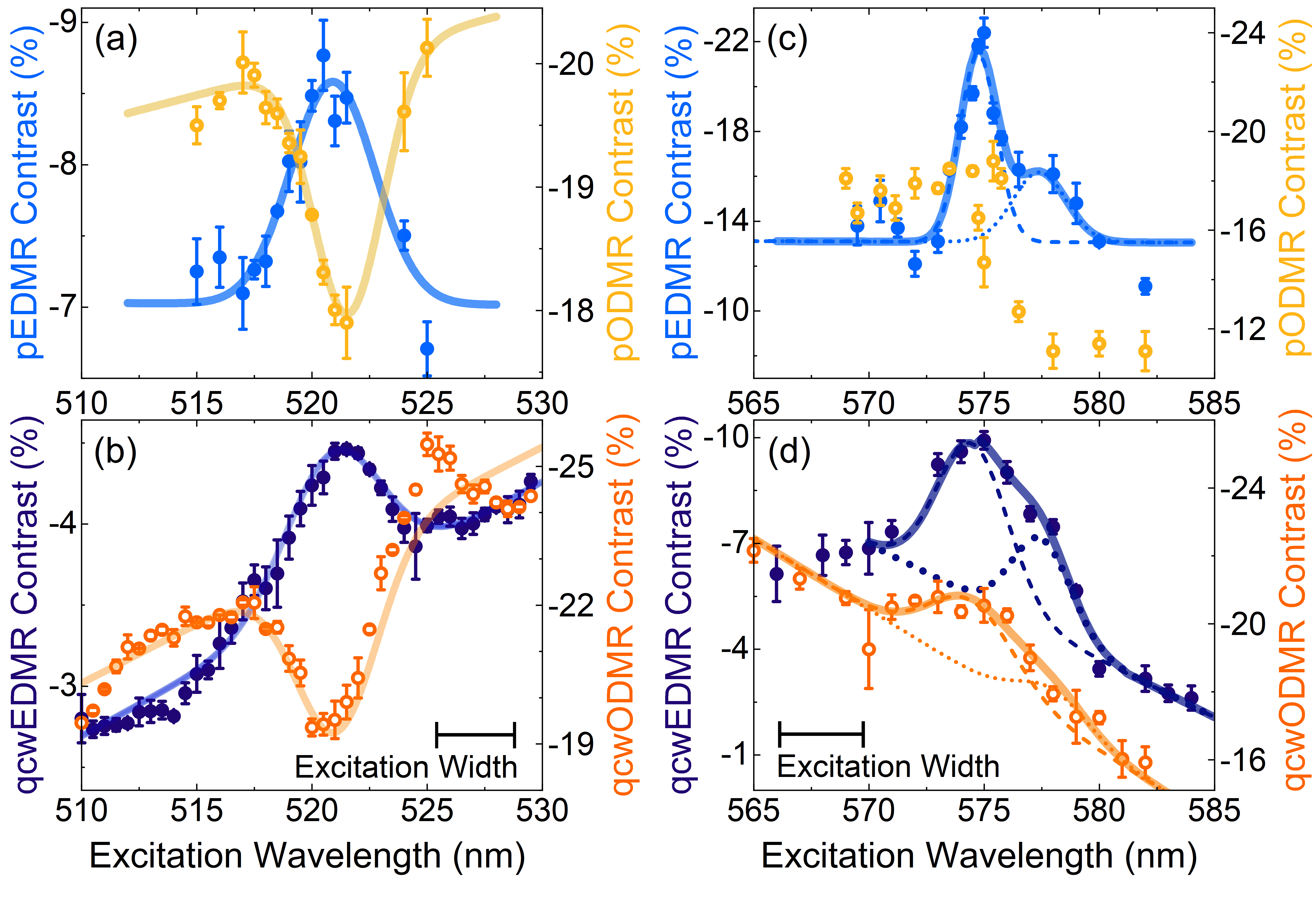}
\caption{\label{Pic3} Pulsed and qcw EDMR and ODMR spectra around 521 and $\SI{575}{\nanometer}$, measured on a second sample and fitted with a single (a,b) or two (c,d) Gaussians on a linear background. (a) Around $\SI{521}{\nanometer}$, the pEDMR (pODMR) spectrum shows a peak (dip) of $\SI{3.8(0.4)}{\nanometer}$ FWHM, attributed to an additional electronic level in the conduction band. (b) The same feature is found under qcw excitation but broadened by the excitation width of the supercontinuum source ($\leq \SI{3}{\nanometer}$, black bar). (c) The pEDMR contrast becomes maximal upon resonant excitation of the NV$^0$ ZPL. Besides the dominant, narrow peak at $\SI{575}{\nanometer}$ (FWHM of $\SI{2.0(0.5)}{\nanometer}$), additional side peaks, here at $\SI{577.4}{\nanometer}$, are observed which we attribute to strain splitting of the ZPL. pODMR shows corresponding minima, coinciding with the dominating decrease in contrast for $\lambda_{\text{ex}} \geq \SI{575}{\nanometer}$. (d) The same behavior is observed in the qcwEDMR, where also the qcwODMR shows a weak local maximum.}
\end{figure}

Especially the pEDMR spectrum shows a rich substructure, of which we investigate two regions (shaded gray in Fig.~\ref{Pic2}) with increased wavelength resolution on a second, nominally identical diamond sample. 
The first region from 510 to $\SI{530}{\nanometer}$ (Fig.~\ref{Pic3}a) exhibits a peak (dip) in the pEDMR (pODMR) contrast at $\lambda_{\text{ex}}=\SI{521}{\nanometer}$. A corresponding feature appears in the PCE (PLE) spectrum in Fig.~\ref{Pic2}a as a pronounced peak (dip). Referring to Beha et al.~\cite{PaperWL-Beha}, we can interpret this as an additional electronic level of NV$^-$ located in the conduction band, $\SI{2.38}{\electronvolt}$ above $^3$E. Such a level could facilitate the ionization from the excited triplet state and increase the spin-dependent photocurrent at the cost of PL intensity. In principle, the same effect could also result from a two-center excitation into the excited electronic level of another defect but would require comparatively high defect densities \cite{PaperWL-2LevelExcitation-S}.\par
The second region investigated in detail encloses the ZPL of NV$^0$ at $\SI{575}{\nanometer}$ (Fig.~\ref{Pic3}c), where we find a maximum in pEDMR contrast of $\SI{-23}{\percent}$ for the second sample. We explain this by an exceptionally effective back-conversion process from NV$^0$ to NV$^-$, which is supported by the peaks observed at $\SI{575}{\nanometer}$ in both the PLE and PCE spectra in Fig.~\ref{Pic2}a.
Consistently, also charge-state sensitive PLE measurements found this specific transition particularly critical within the NV center's charge state conversion cycle \cite{PaperWL-AugerProzess-ZPLExcitation, PaperWL-Beha}. In addition to the dominant, reproducible pEDMR peak at $\SI{575}{\nanometer}$, the feature exhibits a locally varying substructure consisting of minor peaks. These can be shifted by up to $\SI{2.5}{\nanometer}$ towards higher or lower excitation wavelengths in dependence of the position on the diamond surface, presumably due to strain splitting of the NV$^0$ ZPL \cite{PaperWL-StrainSplittingZPL-BehaManson, PaperWL-StrinSplittingOLD-Davies, PaperWL-StrainSplittingZPL-Olivero}. In the pEDMR spectrum in Fig.~\ref{Pic3}c, the main peak has a FWHM of $\SI{2.0(0.5)}{\nanometer}$ and is accompanied by a single side peak at $\SI{577.4}{\nanometer}$. pODMR shows a similar but mirrored structure, i.e. exhibiting minima instead of maxima. These minima coincide with the predominant contrast decrease from $\SI{575}{\nanometer}$ onward, which makes the observation of the pODMR structure challenging. 
In addition, pEDMR and pODMR contrast at $\lambda_{\text{ex}}=\SI{575}{\nanometer}$ both show a particularly strong dependence on the optical excitation power. In the inset of Fig.~\ref{Pic2}b, we quantify this in terms of the ZPL visibility $(C_{\text{575}}-C_{\text{off}})\slash C_{\text{off}}$, which describes the contrast $C_{\text{575}}$ at $\SI{575}{\nanometer}$ in comparison to the average contrast $C_{\text{off}}$ between 565 and $\SI{580}{\nanometer}$. While for pODMR, the visibility is consistently negative (corresponding to a dip at $\lambda_{\text{ex}}=\SI{575}{\nanometer}$), the pEDMR visibility shows a more complex behavior including a sign change and a maximum. At high excitation powers, the visibility decreases to zero for both readout methods, presumably related to the beginning saturation of some transitions.\par
The ZPL of NV$^-$ in turn is found at $\SI{637}{\nanometer}$, where recharging from NV$^0$ is no longer possible, and thus appears only as faint peaks in the PCE and pEDMR spectra of Fig.~\ref{Pic2}. The remaining substructure between 520 and $\SI{580}{\nanometer}$ we ascribe to the interplay of multiple effects. These include phonons \cite{PaperWL_PSBStructure_NV} and the absorption spectra of the individual transitions in the NV charge state conversion cycle, but also the presence of other defects such as N$_{\text{s}}^0$. These defects contribute to the PC either directly or indirectly via recombination processes \cite{PaperWL-B-AcceptorDefects} and become even more critical under qcw excitation.\par
We now demonstrate that the contrast in EDMR and ODMR is subject to different excitation dynamics by performing the readout under qcw conditions, where the illumination is delivered in high-intensity $\SI{50}{\picoseconds}$-long pulses, separated by $\SI{12.8}{\nano\second}$. However, as the excited state lifetimes of NV$^0$ and NV$^-$ as well as the ISC transition time are all between 8 and $\SI{20}{\nano\second}$ \cite{PaperWL-Beha, PaperWL_PulsedChargeStateDynamics-Childress, PaperWL-IrLifetime}, similar to the pulse separation, the qcw scheme imposes boundary conditions on the excitation dynamics. As in standard cw experiments \cite{PaperWL-BReview}, the microwave is applied continuously (cf. inset Fig.~\ref{Pic2}c).
In terms of the optical readout, the overall contrast is slightly higher under qcw conditions, but the general shape of the spectrum in Fig.~\ref{Pic2}c is similar for both excitation schemes. This is quite different for EDMR, where the qcw contrast becomes significantly lower over the entire excitation spectrum, but especially for shorter $\lambda_{\text{ex}}$.
We explain this by recalling that the spin-dependent contribution to PC originates exclusively from two-step ionization, which is hindered by the relaxation of NV$^-$ to its ground state between subsequent pulses.
Instead, the generation of background photocurrent from single-photon ionization of NV and N$_{\text{s}}^0$ becomes dominant under qcw conditions. Since especially the ionization cross-section of N$_{\text{s}}^0$ \cite{PaperWL-RosaNesladek_PhotocurrentSpectroscopyLowE} rises towards shorter wavelengths, also the PC background increases, as can be seen in Fig.~\ref{Pic2}a. This causes the effective redshift of the qcwEDMR spectrum in Fig.~\ref{Pic2}c. Similarly, also Hruby et al.~suspect the ionization of N$_{\text{s}}^0$ to be responsible for increased cwEDMR contrast at $\SI{561}{\nanometer}$ compared to $\SI{532}{\nanometer}$ excitation \cite{PaperWL-B-561nm}.
The optical readout on the other hand benefits from the qcw excitation, since reduced photoionization of $^3$E promotes the spin-dependent radiative decay.
As in the pulsed readout, qcwEDMR (qcwODMR) shows a clear peak (dip) around $\SI{521}{\nanometer}$ (cf.~Fig.~\ref{Pic3}a and c) and also the ZPL of NV$^0$ appears as EDMR maximum regardless of the excitation dynamics (cf.~Fig.~\ref{Pic3}b and d). For ODMR, however, the behavior reverses upon qcw excitation and now peaks at $\SI{575}{\nanometer}$, highlighting the complex power dependence of the resonant excitation of NV$^0$.\par
In summary, we investigated the achievable contrast in the electrical and optical readout of NV centers in diamond as a function of the excitation wavelength under different excitation dynamics. We found that EDMR can be particularly well excited at $\SI{521}{\nanometer}$, which we discuss in terms of a further excited state of NV$^-$ in the conduction band, and at $\SI{575}{\nanometer}$, which corresponds to the ZPL of NV$^0$ and highlights the importance of recharging within the NV charge state conversion cycle. In the IIa CVD diamond samples studied here, we observe a maximum electrical contrast of $\SI{-23}{\percent}$ upon resonantly exciting the NV$^0$ ZPL, which encourages using $\SI{575}{\nanometer}$ instead of the typically applied green spectral region in future EDMR applications and integrated sensors. Also, our results suggest a variety of further fundamental studies, particularly into the effect of background doping. 
On the one hand, single centers in very pure, high-quality diamond should allow us to investigate the isolated NV charge state conversion cycle. In type Ib diamond with high N$_{\text{s}}^0$ concentration on the other hand, the recharging might primarily proceed via the nitrogen donors \cite{PaperWL-Manson-NV-N-Pairs} and thus lose its spin dependence. This is particularly critical with respect to the question whether the spin-dependent photocurrent is dominantly carried by electrons or holes.
Combined with simulations, the high flexibility of pulsing a tunable cw laser can also be used to disentangle the separate transitions involved in the NV charge state conversion cycle by studying photocurrent and contrast as a function of excitation power and pulse length, where preliminary experiments revealed unexpectedly strong $P_{\text{ex}}$ dependence of both PC and EDMR contrast, especially at $\SI{575}{\nanometer}$.
Furthermore, the design of multicolor pulse sequences would allow to control specific transitions of the NV center individually and lead to significantly improved charge state pumping or spin-to-charge conversion schemes \cite{PaperWL_Shields_SpinToCharge, PaperWL_NatureC_SpinToCharge, PaperWL_575Recharging}. 
Combined with increased time resolution in photocurrent measurements, these multicolor sequences could not only greatly improve the electrical readout itself, but also help to ultimately distinguish ionization and recharging steps of the NV center experimentally.\par
This work was supported by BMBF through project epiNV (13N15702), by Bayerisches Staatsministerium für Wissenschaft und Kunst through project IQSense via the Munich Quantum Valley (MQV) and by DFG via the Munich Center for Quantum Science and Technology (MCQST, EXC2111).

\bibliography{NV_Wavelength_Todenhagen}

\end{document}